\documentclass[9pt,twocolumn,twoside]{osajnl}
\usepackage{amssymb,epsfig,setspace,color,widetext}
\graphicspath{{figs/}}
\newcommand{\be}{\begin{equation}}
\newcommand{\ee}{\end{equation}}
\newcommand{\bg}{\begin{equation}}
\newcommand{\eg}{\end{equation}}
\newcommand{\bdm}{\begin{displaymath}}
\newcommand{\edm}{\end{displaymath}}
\newcommand{\bea}{\begin{eqnarray}}
\newcommand{\eea}{\end{eqnarray}}
\newcommand{\beas}{\begin{eqnarray*}}
\newcommand{\eeas}{\end{eqnarray*}}
\newcommand{\ba}{\begin{array}}
\newcommand{\ea}{\end{array}}
\newcommand{\nn}{\nonumber}

\newcommand{\bfg}{\begin{figure}}
\newcommand{\efg}{\end{figure}}

\newcommand{\tl}{\tilde}

\newcommand{\fr}{\frac}

\newtheorem{lm}{Lemma}
\newtheorem{cl}{Corollary}
\newtheorem{df}{Definition}

\newcommand{\blm}{\begin{lm}}
\newcommand{\elm}{\end{lm}}
\newcommand{\bcl}{\begin{cl}}
\newcommand{\ecl}{\end{cl}}
\newcommand{\bdf}{\begin{df}}
\newcommand{\edf}{\end{df}}
\newcommand{\brk}{\begin{rm}}
\newcommand{\erk}{\end{rm}}

\newcommand{\lb}{\label}

\newcommand{\om}{\omega}

\newcommand{\Dt}{\Delta}

\newcommand{\veps}{\varepsilon}

\newcommand{\ld}{\lambda}

\newcommand{\gm}{\gamma}

\newcommand{\sg}{\sigma}

\newcommand{\ct}{\cite}
\newcommand{\rf}{\ref}

\journal{ol}
\setboolean{shortarticle}{true}

\title{An intriguing branching of the maximum position of the absorption cross section in Mie theory explained}

\author[1,*]{Ilia L. Rasskazov}
\author[1]{P. Scott Carney}
\author[2]{Alexander Moroz}

\affil[1]{The Institute of Optics, University of Rochester, Rochester, NY 14627, USA}
\affil[2]{Wave-scattering.com (e-mail: wavescattering@yahoo.com)}
\affil[*]{Corresponding author: irasskaz@ur.rochester.edu}

\begin{abstract}
A potential control over the position of maxima of scattering and absorption cross-sections can be exploited to better tailor nanoparticles for specific light-matter interaction applications. 
Here we explain in detail the mechanism of an appreciable blue shift of the absorption cross-section peak relative to a metal spherical particle localized surface plasmon resonance (LSPR) and remaining scattering and extinction cross sections. 
Such a branching of cross sections maxima requires a certain threshold value of size parameter $x\approx 0.7$ and is a prerequisite for obtaining high fluorescence enhancements, because the spectral region of high radiative rate enhancement becomes separated from the spectral region of high non-radiative rate enhancement.
A consequence is that the maximum of the absorption cross section cannot be used as the definition of the LSPR position for $x\gtrsim 0.7$.
\end{abstract}

\setboolean{displaycopyright}{true}

\begin{document}

\maketitle

\begin{figure}
    \centering
    \includegraphics{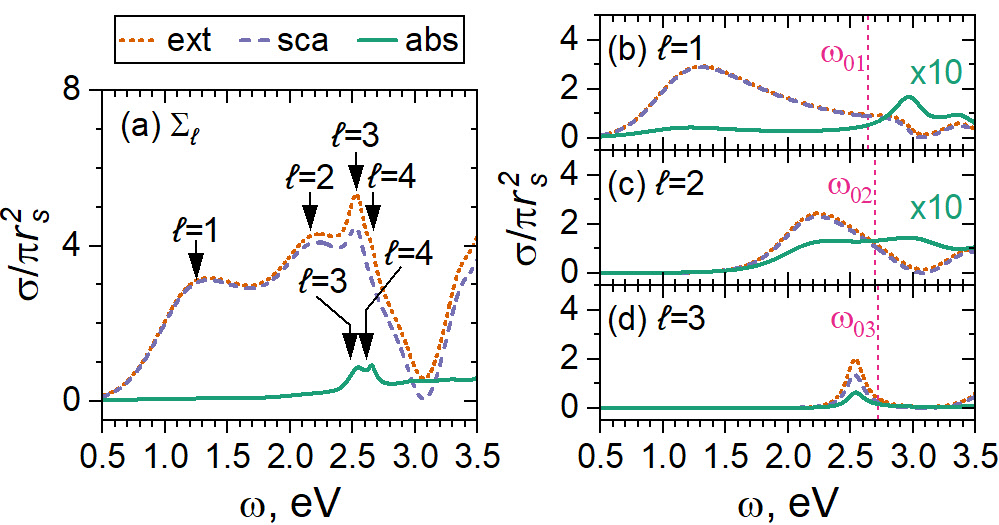}
    \caption{
    (a) Fundamental cross sections of a homogeneous Au particle of radius $r_s=170$~nm calculated via Mie theory.
    Whereas the maxima and minima of the extinction ($\sigma_{ext}$) and scattering ($\sigma_{sca}$) cross-sections occur in unison, the absorption ($\sigma_{abs}$) cross section does not exhibit any maximum at the dipole ($\ell=1$) and quadrupole ($\ell=2$) peaks of $\sigma_{ext}$. 
    The first maximum of $\sigma_{abs}$ is largely blue shifted and occurs nearly at the octupole ($\ell=3$) peak of $\sigma_{ext}$. 
    The panels (b)-(d) show in detail the cross sections in the respective angular-momentum channels.
    The quasi-static LSPR positions $\om_{0\ell}$, defined implicitly by $\veps_s=-(\ell+1)/\ell$, are shown by pink vertical lines. 
    Note the reshuffling of the natural order of the maxima of $\sigma_{abs}$ of the respective multipoles.
    }
    \label{fig:spec}
\end{figure}

The electromagnetic theory of light scattering from small particles with sizes comparable with the wavelength of the incident illumination has the long history~\cite{Mie1908}.
In a number of applications, the control of the absorption is preferable, which led to the extensive theoretical and numerical works in this direction~\cite{Fleury2014,Grigoriev2015,Ladutenko2015,Mirzaei2015,Butet2016a,Miroshnichenko2018,Tribelsky2020},
as well as various experimental applications \ct{Boyer2002,Ni2012,Harris2006,Neumann2013}.
Our recent study \ct{Sun2020} has revealed that optimal metallo-dielectric core-shell particles designed to obtain the highest possible fluorescence enhancements have the spectral region of high radiative rate enhancement {\em red}-shifted and well separated from the {\em blue}-shifted spectral region of high non-radiative rate enhancement \ct[Fig. 4]{Sun2020}. Herein and below, any red or blue shift, unless specified otherwise, refers relative to the particle localized surface plasmon resonance (LSPR). 
For the purpose of this letter, a LSPR will be identified as a maximum of the total (extinction) cross section.
The maximum of radiative rate enhancement coincides with the maximum of near-field (NF) \ct{Sun2020} and the NF peak red shift has been known for a long time~\cite{Messinger1981}. 
Contrary to that, there seem to be no reports of a blue shift of the maximum of the absorption cross 
section, $\sigma_{abs}$, responsible for the maximum of non-radiative rate enhancement \ct{Sun2020}. 
Below we explain in detail the mechanism underlying the puzzling and appreciable, 
yet so far overlooked, blue shift of the absorption cross-section peak.
Because Fig.~\ref{fig:spec} shows that such a pronounced blue shift of the absorption cross section is intrinsic already for a homogeneous spherical metal particle of sufficiently large radius $r_s$, we shall here, for the sake of simplicity, focus only on {\em homogeneous} particles.

The resulting cross sections in the Mie theory are given as an infinite sum over all momentum channels $\ell\ge 1$ and both polarizations~\cite{Newton1982,Bohren1998}.
According to Eqs. (2.135-8)  of Ref.~\cite{Newton1982}, any given angular momentum channel $\ell$ of one of polarizations ($A=E$ for electric (or TM) polarization, and $A=M$ for magnetic (or TE) polarization) contributes the following partial amount to the resulting cross sections shown in Fig. \ref{fig:spec}(a),

\begin{eqnarray}
\sigma_{sca;\ell} &= & \frac{6\pi}{k^2}\, |T_{A\ell}|^2,
\label{sgsc}
\\
\sigma_{abs;\ell} &= & \frac{3\pi}{2k^2}\, \left( 1- |1+2T_{A\ell}|^2\right),
\label{sgabs}
\\
\sigma_{ext;\ell} &= & -\frac{6\pi}{k^2}\, \Re(T_{A\ell}),
\label{sgtot}
\end{eqnarray}
where $T_{A\ell}$ are the T-matrix elements,
$\Re$ takes the real part,
$k=2\pi/\ld$ is the wavenumber, and $\ld$ is the incident wavelength in the host medium.
For $\ell=1$, the above expressions can be easily rephrased in terms of a particle polarizability, $\alpha$, on substituting $2ik^3\alpha/3$ for $T_{E1}$.
In the case of a homogeneous sphere, the respective T-matrix elements in a given $\ell$-th angular momentum channel are~\cite[Eqs. (2.127)]{Newton1982}
\begin{equation}
T_{A\ell} = - \frac{ m [xj_\ell(x)]' j_\ell(x_s)  
                        - j_\ell(x) [x_sj_\ell(x_s)]'}
             {  m [x h_\ell(x)]' j_\ell(x_s)  
               -  h_\ell(x) [x_sj_\ell(x_s)]'}, 
\label{miecoef}
\end{equation}
where $x=k r_s$ is the dimensionless size parameter, 
$r_s$ is the sphere radius, 
$x_s=x\sqrt{\tilde{\veps}_s}$, where $\tilde{\veps}_s=\veps_s/\veps_h$ is the relative dielectric contrast, with $\veps_s$ ($\veps_h$) being the sphere (host) permittivity. In what follows, the host will be assumed to be air ($\veps_h=1$).
One has, assuming nonmagnetic media, $m=1$ for magnetic and 
$m=\tilde{\veps}_s$ for electric polarization,
$j_\ell$ and $h_\ell=h_\ell^{(1)}$ are the conventional spherical functions (see Sec. 10 of Ref. \cite{Abramowitz1973}), and
prime denotes the derivative with respect to the argument.
The Drude model is used throughout this letter for Au permittivity,
\begin{equation}
    \veps = \veps_\infty - \dfrac{\om_p^2}{\om (\om + i \gamma) } ,
\lb{dfit}
\end{equation}
with high-frequency permittivity limit $\veps_\infty = 9.5$, the bulk plasma 
frequency $\om_p = 8.9488$~eV, and damping constant $\gm = 0.06909$~eV, which is
optimized to fit experimental data of Johnson and Christy \cite{Johnson1972}.
The host medium will be assumed to be air characterized by permittivity $\veps_h=1$.

\begin{figure}[t!]
    \centering
    \includegraphics{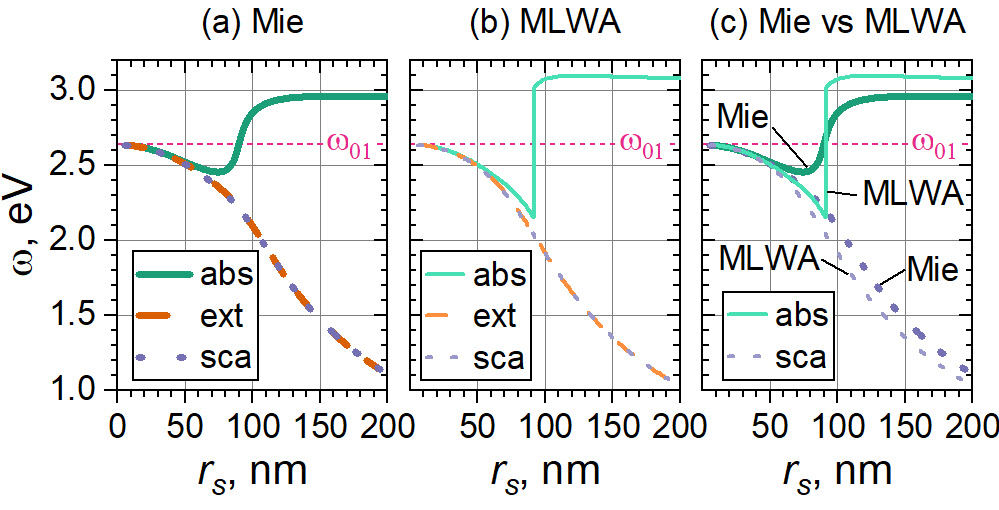}
    \caption{Evolution of the spectral positions of the maxima of absorption, extinction and scattering cross sections with increasing sphere radius $r_s$ of Au sphere in air for dipole $\ell=1$ electric mode contribution calculated by (a) exact Mie theory, and (b) on using \eqref{sgabsdm}~--~\eqref{sgtotsdm} of MLWA. 
    Note a very nice agreement between the respective results (c). 
    Whereas the scattering and extinction cross section peak positions continue in their synchronized red shift relative to $\om_{01}$ with increasing $r_s$, the absorption cross-section peak reverses its initial red shift to an appreciable blue shift which remains essentially stable for $r_s\gtrsim 100$ nm. 
    Note in passing that a spectral gap between the LSPR (determined as the extinction cross section peak position) and the absorption cross-section peak position can be as large as 2 eV for $r_s\approx 200$ nm.}
  \label{fig:MieMLWAl1}
\end{figure}

As seen in Fig. \ref{fig:MieMLWAl1}(a), with increasing particle radius up to $r_s\simeq 70$ nm, the maxima of all cross sections of the dipole term of the Mie series experience initially monotonically increasing {\em red} shift from the initial quasi-static LSPR position at $\om_{01}$. 
Thereafter the maximum of the absorption cross section (\rf{sgabs}) gradually reverses its red shift relative to $\om_{01}$ into an increasing blue shift, which for $r_s\approx 100$ nm can be as large as $100$ nm. 
The position of the maximum of the absorption cross section becomes eventually nearly constant with increasing size parameter $x$ in a {\em blue} shifted region (centered around the frequency implicitly given by \eqref{vepsza} below) relative to $\om_{01}$.

\begin{figure}
    \centering
    \includegraphics{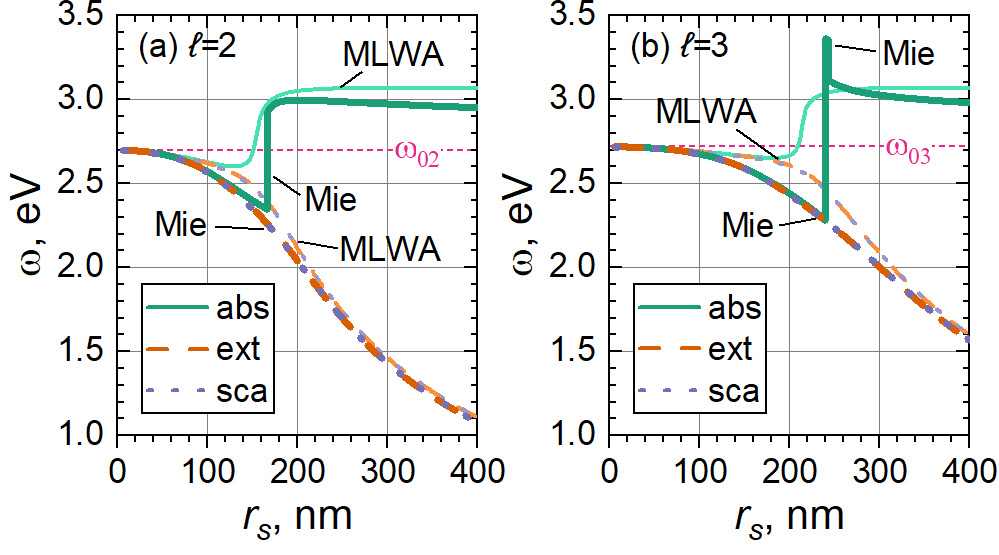}
    \caption{Evolution of the spectral positions of the maxima of quadrupole ($\ell=2$) and octupole ($\ell=3$) absorption, extinction and scattering cross sections with increasing sphere radius $r_s$.}
    \label{fig:MieMLWAgl1}
\end{figure}

As suggested by Figs. \ref{fig:MieMLWAl1}(c) and \ref{fig:MieMLWAgl1}, the modified long-wavelength approximation (MLWA) agrees very well with the exact Mie theory over entire size range up to $x\approx 5$ not only for dipole, but also for higher multipole contributions. 
Therefore, the essentials of the blue shift can be captured by the MLWA. 
The usual MLWA \ct{Meier1983,Kelly2003,Moroz2009} is a limiting form of the Mie dipole term, that, unlike the usual quasi-static Rayleigh approximation, keeps both dynamic depolarization ($\sim x^2$) and radiative reaction ($\sim x^3$) terms. 
This is why the MLWA can account for a size-dependent red shift of the dipole LSPR, whereas the Rayleigh approximation cannot. 
Here we use the following MLWA form of the $T$-matrix which is valid in any given channel $\ell$ (cf. Eq. (A3) of \cite{Moroz2009} for $\ell=1$),
\bea
T_{E\ell} &\sim& \fr{i R(x)} {F + D(x) -i R(x)},
 \lb{tmtflr3c}
\nn\\
F &=& \tilde{\veps}_s + \fr{\ell+1}{\ell},
 \nn\\
D(x) &=& 
 \left(\fr{\ell-2}{\ell+1}\, \tilde{\veps}_s + 1\right) 
\fr{(\ell+1)(2\ell+1)}{\ell(2\ell-1)(2\ell+3)}\, x^2 ,
 \nn\\
R(x) &=&  
\fr{\ell+1}{\ell (2\ell-1)!! (2\ell+1)!! }\, (\tilde{\veps}_s-1)\, x^{2\ell+1}\cdot
\eea
\eqref{tmtflr3c} is derived by using asymptotic expressions for spherical Bessel and Hankel 
functions used in (\rf{miecoef}).
It makes transparent that $T_{E\ell}$ in any given channel is determined solely by a size independent quasi-static Fr\"ohlich term $F$, a {\em dynamic depolarization} term $D$ ($\sim x^2$), and a {\em radiative reaction} term $R$ ($\sim x^{2\ell+1}$).
The vanishing of the size independent $F$ in the denominator yields the usual quasi-static Fr\"ohlich LSPR condition, which determines the quasi-static LSPR frequencies $\om_{0\ell}$. 
In the case of Drude fit (\rf{dfit}) of $\veps_s$ one finds $\om_{0\ell}=\om_p/\sqrt{\veps_\infty+[(\ell+1)\veps_h/\ell]}$. 
One refers, somewhat misleadingly, to a {\em unitarity}, if the substitution of a given approximation to $T_{E\ell}$ into the above equations yields $\sigma_{ext;\ell}=\sigma_{sca;\ell}+\sigma_{abs;\ell}$. 
The MLWA can be shown to satisfy unitarity. 
In contrast, the usual {\em Rayleigh} limit, which amounts to setting $D(x)=R(x)\equiv 0$ in the denominator of the dipole MLWA $T_{E1}$ in (\rf{tmtflr3c}), yields a {\em purely imaginary} $T_{E\ell}$ for real $\veps_s$ (i.e., a {\em purely real} polarizability) and violates the unitarity \ct{Moroz2009}. 

In what follows we shall focus on the dipole MLWA. 
The higher order multipole MLWA can be treated similarly. 
On substituting \eqref{tmtflr3c} into \eqref{sgsc} -- \eqref{sgtot}, one finds the following cross sections of the dipole MLWA contribution:
\bea
\sg_{abs;1} &=& \fr{4\pi}{15 k^2}\, \frac{9\, x^3 \left(x^2+5\right) \Im(\tl\veps_s)}{\left|
\tl\veps_s+2-\fr{3}{5} (\tl\veps_s-2) x^2-i \fr{2}{3} (\tl\veps_s-1) x^3\right|^2},
\lb{sgabsdm}
\\
\sg_{sca;1} &=& \fr{4\pi}{15 k^2}\, \frac{10\, x^6 \left| \tl\veps_s-1\right|^2}{\left|\tl\veps_s+2-\fr{3}{5} (\tl\veps_s-2) x^2-i \fr{2}{3} (\tl\veps_s-1) x^3\right|^2},
\lb{sgscsdm}
\\
\sg_{ext;1} &=& \fr{4\pi}{15 k^2}\, \frac{9\, x^3 \left(x^2+5\right) \Im(\tl\veps_s)+10\, x^6 \left|\tl\veps_s-1\right|^2}
{\left|\tl\veps_s+2-\fr{3}{5} (\tl\veps_s-2) x^2-i \fr{2}{3} (\tl\veps_s-1) x^3\right|^2},
\lb{sgtotsdm}
\eea
where $\Im(\tl\veps_s)$ denotes the imaginary part of $\tl\veps_s$.
For $\Im(\tl\veps_s)=0$, the common denominator $|\Dt|^2$ of the dipole MLWA cross-sections (\rf{sgabsdm})-(\rf{sgtotsdm}) vanishes at
\bg
\tl\veps_s \approx -2-\fr{12 x^2}{5}   
\lb{vepsz}
\eg
up to the order $x^3$, in which case $\Dt\approx {\cal O} (x^3)$.
For $\Im(\tl\veps_s)\ne 0$, one can approximate $|\Dt|^2$ as
\bg
|\Dt|^2\approx \left|\left(1- \fr{3x^2}{5} \right)\Im(\tl\veps_s) + 2x^3\right|^2 
+
\left|
\fr{2x^3}{3}\, \Im(\tl\veps_s)\right|^2.
\nn   
\eg
Eq.  (\rf{vepsz}) imposes restriction on the real part of $\tl\veps_s$ \ct[Eq. (B1)]{Moroz2009} explaining the observed initial size-dependent red shift of all cross sections (Fig.~\ref{fig:MieMLWAl1}(a)).

\begin{figure}
    \centering
    \includegraphics{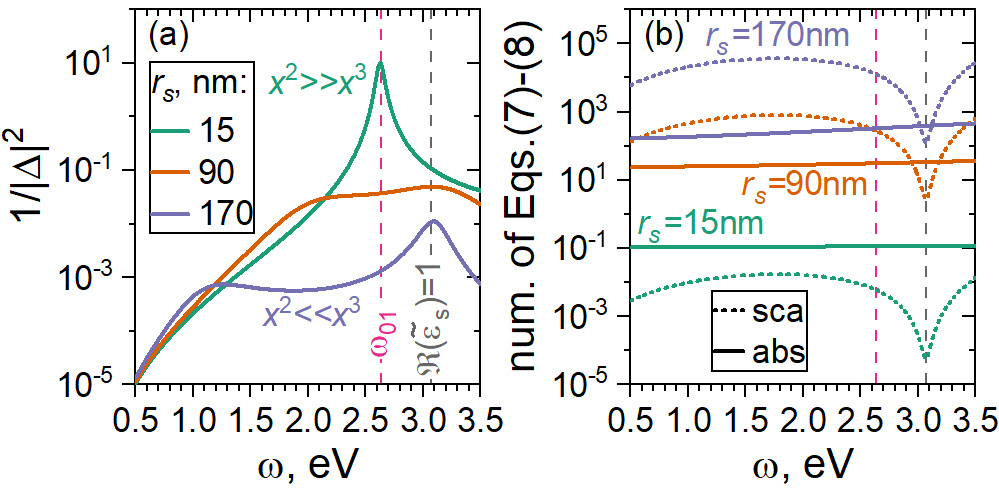}
    \caption{
    Evolution of the blue-shifted maximum of the dipole $\sg_{abs;1}$ with increasing $x$.
    (a) The minimum of denominator and (b) a corresponding value of the numerator for absorption and scattering cross sections of \eqref{sgabsdm} and \eqref{sgscsdm}.
    A double peak of $1/|\Dt|^2$ for $r_s=170$ nm, with the maximum at the frequency implicitly given by \eqref{vepsza}, can be clearly identified.
    Note that the numerator of $\sg_{sca;1}$ is smaller than that of $\sg_{abs;1}$  at (\ref{vepsza}).
    Panels (a) and (b) also illustrate the well-known fact why smaller particles have much larger absorption than larger particles~\cite{Bohren1998}.}
    \label{fig:frac}
\end{figure}

At the MLWA branching point at $r_s\approx 90$ nm in Fig.~\ref{fig:MieMLWAl1}(b), where the peak position of $\sg_{abs;1}$ begins to deviate from the remaining cross sections, the size parameter $x\approx 0.975$
(cf. $x\approx 0.8$ and the internal size parameter $x_s=\sqrt{\tl\veps_s} x_s \approx 1.1$ for the branching point at $r_s\approx 70$ in the Mie theory in Fig. \ref{fig:MieMLWAl1}(a)). 
The appearance of the branching point and ensuing blue shift of the absorption cross section (\rf{sgabs}) can be explained within the dipole MLWA as follows:
\begin{itemize}

\item[(i)] For $x^2\gg x^3$ the maxima positions are governed by the first two terms of the complex root of $\Dt$ on the rhs of \eqref{vepsz}, as was already observed by Bohren and Huffman~\cite[Sec. 12.1.1]{Bohren1998}, yielding the red-shifted behaviour of the maximum of $1/\Dt$ relative to $\om_{01}$ with increasing $x$. 
The latter is the chief cause of the synchronized red-shifted behaviour of the maxima of all three fundamental scattering cross-section relative to $\om_{01}$ with increasing $x$ up to $x\approx 1$. 
This is why, unlike the Rayleigh approximation, the MLWA can account for the size-dependent red shift of the dipole LSPR.

\item[(ii)] As soon as $x^3\sim x^2$, the $x^3$-term of $\Dt$ in Eqs.~(\rf{sgabsdm}) -- (\rf{sgtotsdm}) begins to dominate. 
One can achieve another maximum of $1/\Dt$ if the $x^3$-term is rendered as small as possible, 
i.e. $\tl\veps_s$ has to be ideally $+1$. The latter can, in the case of dissipative
media (Au), be in principle achieved only with a suitably tailored gain.
In the present case we are left with the condition 
\bg
\Re(\tl\veps_{max;abs}) \approx 1,
\lb{vepsza}
\eg
which means that the blue-shifted maximum of $\sg_{abs;1}$ coincides with the condition of minimizing the radiative reaction (Fig. \rf{fig:frac}(a)). This causes a maximum of $\sg_{abs;1}$ but not of $\sg_{sca;1}$ ((Fig. \ref{fig:spec}). 
The reason why the other maximum of $1/\Dt$ does not cause a maximum of $\sg_{sca;1}$ is that the latter has 
$|\tl\veps_s-1|$ in its numerator, which becomes very small, in contrast to 
$\sg_{abs;1}$ having $\Im(\tl\veps_s)$ in its numerator (Fig. \rf{fig:frac}(b)).
Note, $\Re(\veps_{max;abs}) \approx -1.5$ for $r_s\approx 91.5$~nm, and $\Re(\veps_{max;abs})\approx 1$ for $r_s\gtrsim 107$~nm.

\end{itemize}
The condition (\ref{vepsza}) explains why the position of the blue-shifted maximum of $\sg_{abs;1}$ remains substantially constant with increasing $r_s$, soon after it branched off from the position of maxima of other two cross-sections. 
With the parameters of the Drude fit (\rf{dfit}), the zero of $\Re(\tl\veps_s)$ occurs at $\om_z=\om_p/\sqrt{\veps_\infty}= \om_p/\sqrt{9.5}\approx 2.99144$ eV. 
The $\Re(\tl\veps_s)=1$ occurs at $\om_{rc}= \om_p/\sqrt{8.5}\approx 3.0694$ eV. 
One can clearly observe in Figs.~\ref{fig:MieMLWAl1} and \ref{fig:frac}(a) how the blue-shifted maximum of $\sg_{abs;1}$ stabilizes around $\om_{rc}$ with increasing $x$. 
Obviously, the blue shift of absorption is possible only if $\Im(\tl\veps_s)$ at the frequency implicitly given by \eqref{vepsza} is sufficiently small, as in the present case of Au. 
Provided that $\Im(\tl\veps_s)$ at (\rf{vepsza}) is sufficiently large, this may prevent $\Dt$ from acquiring a local minimum and the blue shift of absorption may be absent. 
The situation for $\ell>1$ is analogous to that for $\ell=1$ in that the radiative reaction term $R$ in \eqref{tmtflr3c} maintains its factor $\tl\veps_s-1$. 
A slight modification to $\ell=1$ case results from that the $\ell$-dependent factor $(\ell+1)/[\ell (2\ell-1)!! (2\ell+1)!!]$ of $R(x)$ in \eqref{tmtflr3c} {\em decreases} much faster with increasing $\ell$ than the $\ell$-dependent factor $(\ell+1)(2\ell+1)/[\ell(2\ell-1)(2\ell+3)]$ of $D(x)$.
Therefore, the threshold size parameter $x$ value required for the MLWA $\sg_{abs;\ell}$ to exhibit a blue-shifted maximum necessarily {\em increases} with increasing $\ell$. 
Again, this is clearly seen in Fig.~\ref{fig:MieMLWAgl1}. 
Consequently, the build-up of the blue-shifted maxima of $\sg_{abs;\ell}$ is {\em not} uniform in $\ell$. 
For a given fixed $x$, only the lowest multipoles have blue-shifted maximum, whereas the maxima of the remaining multipoles continue to be red-shifted. The latter causes a rearrangement of the natural order of the $\ell$-pole absorption maxima: the lowest absorption peak of $\sg_{abs}$ may be due to $\ell=3$ followed by $\ell=4$, and only later the absorption maxima of $\ell=1,2$ appear (Fig. \ref{fig:spec}).
Overall, the above condition (\ref{vepsza}) defines a rare location at which the total $\sg_{abs}$ ($\gg\sg_{sca}$) is the dominant contribution to $\sg_{ext}$ and the single-scattering albedo (the ratio of scattering efficiency to total extinction efficiency) acquires its minimum (Fig. \ref{fig:spec}(a)).

To conclude, the properties of small metal particles continue to surprise.
Whenever one thinks a full understanding has been reached, some unexpected connection, or an unnoticed property appears. 
We have demonstrated that at the size parameter value $x\approx 0.7$ the maximum of the absorption cross section $\sg_{abs;1}$ can be appreciably blue shifted relative to the quasi-static position $\om_{01}$ of the dipole LSPR.
An obvious consequence is that the maximum of $\sg_{abs;1}$ can no longer be used as the definition of the LSPR position for $x\gtrsim  0.7$. 
A threshold size parameter value required to exhibit a blue-shifted maximum of $\sg_{abs;\ell}$ for an $\ell$-pole increases with increasing $\ell$. 
The latter causes a rearrangement of the natural order of $\ell$-pole absorption maxima (e.g. the lowest absorption peak of $\sg_{abs}$ may be due to octupole ($\ell=3$) as shown in Fig. \ref{fig:spec}(a)).
The blue shift of absorption is possible only if $\Im(\tl\veps_s)$ at the frequency implicitly given by \eqref{vepsza} is sufficiently small, as in the present case of Au nanosphere. 
Our results bring us one step closer to an ideal world scenario, where one would be able to achieve a control over the position of maxima of basic cross-sections in order to better tailor nanoparticles to specific light-matter interaction applications. 
They could be of immediate interest not only for enhanced fluorescence \ct{Sun2020}, 
but also for nonlinear optics~\cite{Butet2016a}, surface enhanced Raman spectroscopy~\cite{VanDijk2013,Sivapalan2013}, heat management, thermophotovoltaic, photothermal imaging \ct{Boyer2002}, thermoplasmonics, such as thermally enhanced surface-chemistry \ct{Ni2012}, plasmonic heating \ct{Harris2006}, optoplasmonic evaporation, or solar vapor generation enabled by nanoparticles \ct{Neumann2013}.

It is worthwhile to mention that recent work~\cite{Yezekyan2020} deals with a related issue within the electrostatic approximation (i.e. the Rayleigh limit), without any use of MLWA, and without any comparison with the exact Mie results. 
Under these limitations, it is difficult to argue for a blue shift if already the well-known size-dependent red shift (\rf{vepsz}) relative to $\om_{0\ell}$ cannot be properly accounted for.



\end{document}